\documentclass[twocolumn,pre,10pt,aps,floatfix]{revtex4-1}
\usepackage[english]{babel}
\usepackage{graphicx}
\usepackage{amsmath}
\usepackage{nccmath}
\usepackage{mathtools}

\DeclarePairedDelimiterX\braket[2]{\langle}{\rangle}{#1\,\delimsize\vert\,\mathopen{}#2}

\usepackage[colorlinks,linkcolor=blue,citecolor=blue,urlcolor=blue]{hyperref}
\usepackage{color}
\usepackage{soul}
\definecolor{dblue} {RGB}{28,130,185}
\let\oldaddcontentsline\addcontentsline
\newcommand{\stoptocentries}{\renewcommand{\addcontentsline}[3]{}}
\newcommand{\starttocentries}{\let\addcontentsline\oldaddcontentsline}
\definecolor{nred}{RGB}{224,0,0}
\definecolor{darkgreen}{rgb}{0,0.60,.2}
\definecolor{npink}{RGB}{255,0,255}

\begin{document}
%\title{Charge transport in tilted chains with various two-particle interactions}
\title{Dependence of the charge transport in tilted chains on the choice of two-body interaction}
\author{Bartosz Krajewski}
\affiliation{Institute of Theoretical Physics, Faculty of Fundamental Problems of Technology, \\ Wroc\l aw University of Science and Technology, 50-370 Wroc\l aw, Poland}
\author{Marcin Mierzejewski}
\affiliation{Institute of Theoretical Physics, Faculty of Fundamental Problems of Technology, \\ Wroc\l aw University of Science and Technology, 50-370 Wroc\l aw, Poland}
\date{\today}
\begin{abstract}
We study tilted chains of spinless fermions in the presence of the nearest-neighbor density-density interaction for which the noninteracting counterpart displays Stark localization. We demonstrate that the latter two-body interaction can be decomposed into two (orthogonal) parts which, respectively, commute and do not commute with the single-particle Hamiltonian. We derive an explicit form of the noncommuting part that decreases with tilt and
describes the nearest-neighbor correlated hopping and the pair-hopping interaction. When the density-density coupling is replaced by the pair-hopping interaction of the same magnitude then the charge dynamics may be faster by a few orders of magnitude than in the original model.
\end{abstract}
\maketitle
\stoptocentries
%================================================================================

\section{Introduction}

The study of many-body localization (MBL) has significantly advanced our understanding of non-equilibrium dynamics in quantum systems.  MBL was introduced in systems with disordered potentials, where it was conjectured that despite the presence of interactions, one can still observe the breakdown of thermalization, resulting in localized eigenstates~\cite{basko06, gornyi05, oganesyan07, Rahul15, altman15, alet_laflorencie_18, abanin2019}. This phenomenon has been extensively studied both theoretically~\cite{pal10,barisic10,luitz15,torres15,Devakul2015,Hauschild_2016,bertrand_garcia_16,suntajs_bonca_20a, suntajs_2020, sierant_lewenstein_20,sierant2020} and experimentally~\cite{schreiber15,smith2016,luschen17,lukin2019,leonard2023}, primarily in one-dimensional settings. It is characterized, among others, by extremely slow dynamics~\cite{znidaric08,serbyn13_1,mierzejewski2016,luschen17,bordia2017_1,serbyn2017,Bera2019,chanda2020}, logarithmic growth of entanglement entropy in time~\cite{znidaric08,bardarson12,kjall14,baygan15, pietracaprina_parisi_17}, and subdiffusive transport for weaker disorders~\cite{berkelbach10, lev15, barisic16, steinigeweg16, bera2017, prelovsek116}.
Although finite strongly disordered systems appear localized, the fate of localization in macroscopic systems is still under debate~\cite{abanin2019,Panda2020,kieferemmanouilidis_unanyan_20, sels2020, kieferemmanouilidis_unanyan_21,abanin_bardarson_21, leblond_sels_21, vidmar2021, Sels_dilute_2021,Sels_2022,sierant2022,evers2023}.

Similar research may also be carried out beyond the framework of disorder-induced localization to investigate whether non-ergodic behavior can arise in interacting systems without quenched disorder. One particularly intriguing direction concerns the Stark many-body localization that may exist in tilted systems subject to linear potential gradients~\cite{schulz2019,nieuwenburg2019,taylor2020,yao2021,yao2021b,falcao2023}. These systems are accessible experimentally in cold atoms and trapped ions experiments~\cite{sanchez2020,scherg2021,kohlert2023} and offer a unique way to study the ergodicity breaking transition. The suppression of transport in tilted systems \cite{nandy2024} and disordered systems \cite{prelov2023_dif} appears similar
in that the transport coefficients decrease exponentially with either the strength of disorder or with the strength of the tilt. However, the origins of localization in the non-interacting limits of both models are different. In particular, the tilted model of noninteracting particles exhibits the Wannier-Stark localization that can be linked to energy conservation~\cite{stark1962}. Interacting Stark chains conserve the dipole moment, implying that transport at small wave-vectors is subdiffusive~\cite{moudgalya2020, scherg2021, nandy2024}.

It has recently been argued that the density-density interaction commonly used to study MBL is not a truly representative example of the many-body interaction
since only a small part of this two-body interaction acts as a perturbation to the Anderson insulator~\cite{krajewski_vidmar_22}. Specifically, the most of the density-density interaction can be expressed in terms of occupations of the single-particle Anderson states, meaning that this part does not disturb the Anderson insulator. The remaining part of the interaction represents the true perturbation that, however, strongly decreases with the disorder strength. Eventually, for strong disorders the true perturbation may be too small to be accurately detected by numerical simulations of finite systems, thus resulting in contradictory conclusions. To illustrate this problem, we have studied models for which the strength of the true perturbation is of the same order of magnitude as the single-particle Hamiltonian~\cite{krajewski_vidmar_22} or at least does not decrease with disorder~\cite{krajewski2023}. In such models, the indicators of localization are strongly suppressed,
and the finite-size scaling suggests ergodicity of macroscopic systems.   

The very same problem emerges also in the case of the Stark MBL. However, in contrast to disordered systems, the single-particle eigenstates in the Stark chains have a simple analytical form which allows one to study tilted systems analytically~\cite{nieuwenburg2019}. Taking advantage of this property, in this work we calculate analytically the form of the true perturbation for a tilted model with density-density interaction up to the second order in the high-field expansion,  $1/F$. We show that the true perturbation has a form of correlated hopping and pair-hopping interaction on the neighboring lattice sites. Finally, we demonstrate that the slow dynamics observed in a chain with density-density interaction originates,  at least partially, from the smallness of the true perturbation. In particular, when the density-density interaction is replaced by the pair hopping interaction of the same strength, then the dynamics is shown to speed up by a few orders of magnitude.

The paper is organized as follows. In Sec.~\ref{sec:true_pert} we recall the notion of the true perturbation and the method of its calculation. We calculate the norm of the true perturbation for tilted model in Sec.~\ref{subsec:norm} and calculate its analytical form in Sec.~\ref{subsec:form}. In Sec.~\ref{sec:dynamics} we present numerical results of charge dynamics for tilted model subject to various types of interaction chosen based on form of the true perturbation. We summarize and discuss our results in Sec.~\ref{sec:summary}

\section{True perturbation in the Stark model with density-density interaction}
\label{sec:true_pert}

\subsection{The strength of the true perturbation}
\label{subsec:norm}

We consider an interacting tilted chain of length $L$ with open boundary conditions. The system is described by the Hamiltonian $H=H_0+H_V$. The first part, $H_0$, is a single-particle model that exhibits the Wannier-Stark localization
% \begin{equation}
%     H_0 = \sum_j \left(c_j^\dagger c_{j+1} + H.c.\right) +F\sum_j j\left(n_j-\frac{1}{2}\right),
% \end{equation}
\begin{equation}
    H_0 = \sum_j \left(c_i^\dagger c_{i+1} + H.c.\right) +F\sum_j j \tilde{n}_j.
\end{equation}
Here, $c^\dagger_j$ creates a spinless fermion at site $j$, $n_j=c_j^\dagger c_j$, $\tilde{n}_j=n_j-\frac{1}{2}$ and $F$ is a tilt of the lattice. The second term, $H_V$, is a density-density interaction 
% \begin{align}
%     H_V&=V\sum_j h(j),\\
%     h(j)&=\left(n_j-\frac{1}{2}\right)\left(n_{j+1}-\frac{1}{2}\right) \label{h_dd}
% \end{align}
\begin{align}
    H_V&=V\sum_j h(j), \quad \quad
    h(j)=\tilde{n}_j\tilde{n}_{j+1}. \label{h_dd}
\end{align}
The main focus of this work is to decompose $H_V$ into two
parts $H_V=H^{\parallel}_V +H^{\perp}_V$ such that $[H^{\parallel}_V,H_0]=0$. It is rather obvious that only $H^{\perp}_V$ may perturb the Wannier-Stark localization; thus we call this term the
true perturbation. First, we determine how the strength of
$H^{\perp}_V$ depends on $F$, while its explicit form will be discussed in subsequent subsection.

The single-particle part of the Hamiltonian may be written in a diagonal form~\cite{nieuwenburg2019}
\begin{equation}
    H_0=\frac{1}{2}\sum_\alpha \varepsilon_\alpha Q_\alpha+\mathrm{const},
\end{equation}
with
\begin{equation}
    Q_\alpha=2f_\alpha^\dagger f_\alpha-1,\quad f^{\dagger}_\alpha = \sum_j 
    \langle j | \alpha \rangle c^{\dagger}_j.
    \label{Q_alph}
\end{equation}
Away from the boundaries of the studied chain, the single-particle energies are equidistant, $\varepsilon_\alpha=F\alpha$, and the single-particle eigenstates   can be expressed by the Bessel functions of the first kind, $\langle j | \alpha \rangle=\mathcal{J}_{j-\alpha}(2/F)$ for which we use $1/F$ expansion \cite{abramowitz1964}
\begin{equation}
    J_{j-\alpha}(2/F)=\sum_{m=0}^\infty\frac{(-1)^m}{m!(m+j-\alpha)!}\left(\frac{1}{F} \right)^{2m+j-\alpha}.\label{J_exp}
\end{equation}
One observes that the wave function $\langle j | \alpha \rangle$  is Wannier-Stark localized at site $j_0=\alpha$ and exhibits approximately exponential decay in the real space,  $|\langle j | \alpha \rangle| \sim F^{-|j-j_0|}$.

The occupations of the Wannier-Stark states, $Q_\alpha$, as well as their products, $Q_{\alpha,d}^{(2)}\equiv Q_\alpha Q_{\alpha+d}$, commute with $H_0$. Therefore, the true perturbation refers to the part of $H_V$ that cannot be expressed by either  $Q_\alpha$ or $Q_{\alpha,d}^{(2)}$. In order to single out the true perturbation we note that $Q_\alpha$ are orthonormal with orthogonality and normalization defined via the (Hilbert-Schmidt)  product
\begin{equation}
\langle  Q_\alpha Q_\beta \rangle=\frac{1}{Z}\mathrm{Tr}(Q_\alpha Q_\beta )=\delta_{\alpha\beta}.
\end{equation}
Here, the trace is carried out over the many-body Hilbert space of dimension $Z$ so that $\langle ... \rangle$ coincides with the grand-canonical averaging at infinite temperature. It is straightforward to show that the products of occupations are also orthonormal, $\langle Q_{\alpha,d}^{(2)}Q_{\alpha',d'}^{(2)}\rangle=\delta_{\alpha\alpha'}\delta_{dd'}$, provided one takes only positive (or only negative) $d$. From now on, we choose $d>0$.

Following the same reasoning as for the disordered systems ~\cite{krajewski_vidmar_22}, we first
determine projections of the interaction term, Eq. (\ref{h_dd}), on $Q_{\alpha,d}^{(2)}$, 
\begin{align}
      H_V^{\parallel} &= V \sum_j h^\parallel(j) \\
    h^\parallel(j) &= \sum_{\alpha,d}\left\langle h(j) Q^{(2)}_{\alpha,d} \right\rangle Q^{(2)}_{\alpha,d}. \label{h_dd_par}
\end{align}
Obviously, all $h^\parallel(j)$ commute with $H_0$ so that the entire $ H_V^{\parallel}$ commutes with the single-particle Hamiltonian. Consequently, the true perturbation is defined as the difference
\begin{align}
      H_V^{\perp} &= H_V-H_V^{\parallel}=V \sum_j h^\perp(j), \\
    h^\perp(j) &=  h(j) - h^\parallel(j) \label{h_dd_perp}.
\end{align}

The most challenging task related with these calculations is to determine the projections in Eq. (\ref{h_dd_par}). In the case of disordered systems, analogous projections have been estimated from numerical calculations ~\cite{krajewski_vidmar_22}. Here, using the single-particle eigenstates from Eq. (\ref{J_exp}), one can obtain analytically the leading contributions to $H_V^{\perp}$.     
For clarity, from now on, we drop the argument of the Bessel function $\mathcal{J}_j(2/F)\rightarrow\mathcal{J}_j$ 
and write the explicit form of the projection

\begin{align}
   \begin{split}
       \MoveEqLeft[2.0]\left\langle h(j) Q^{(2)}_{\alpha,d} \right\rangle 
   = \frac{1}{4}\sum_{k,l,m,n}\mathcal{J}_{k-\alpha}\mathcal{J}_{l-\alpha}\mathcal{J}_{m-(\alpha+d)}\mathcal{J}_{n-(\alpha+d)}\\
   &   \left\langle(2n_j-1)(2n_{j+1}-1)(2c^\dagger_k c_l-\delta_{kl})(2c^\dagger_{m} c_{n}-\delta_{mn}) \right\rangle.
   \end{split}
   \label{proj}
\end{align}
The only nonzero elements in the four-fold sum in Eq.~(\ref{proj}) are those in which the indices of the creation operators $(k,m)$ and the indices of the annihilation operators $(l,n)$ are permutations of $(j,j+1)$. It leaves one with four elements in the sum which can be written in a compact form
\begin{equation}
    \left\langle h(j) Q^{(2)}_{\alpha,d} \right\rangle=\frac{1}{4}\!\left( \mathcal{J}_{j-\alpha}\mathcal{J}_{j-\alpha-d+1}\!-\!\mathcal{J}_{j-\alpha+1}\mathcal{J}_{j-\alpha-d} \right)^2
    \label{proj1}
\end{equation}

The Bessel functions decay for large $F$ as \mbox{ $|\mathcal{J}_m|\sim \left(\frac{1}{F}\right)^{|m|}$}, see also Eq.~(\ref{J_exp}). One may check (for $d>0$) that the indexes of the Bessel functions in Eq. (\ref{proj1}) satisfy the
inequality $|j-\alpha|+|j-\alpha-d+1|\le |j-\alpha+1|+|j-\alpha-d|$,
so that the largest projection occurs for $\alpha=j$ and
$d=1$ when the left-hand side of the latter inequality is minimal. The largest projection reads
\begin{equation}
    \left\langle h(j) Q^{(2)}_{j,1} \right\rangle=\frac{1}{4}\!\left( \mathcal{J}_{0}\mathcal{J}_{0}\!-\!\mathcal{J}_{1}\mathcal{J}_{-1} \right)^2.
    \label{proj2}
\end{equation}
 Then it is straightforward to calculate the squared norm of $||h^\parallel(j)||^2=\langle  h^\parallel(j) h^\parallel(j) \rangle $. The projection in Eq. (13) is the only term that contributes to that norm up to the second order in $1/F$. Using the fact that $Q^{(2)}_{\alpha,d}$ are orthonormal, one finds from Eq. (\ref{h_dd_par})
\begin{align}
    ||h^\parallel (j)||^2&\!=\sum_{\alpha,d}\! 
    \left\langle h(j) Q^{(2)}_{\alpha,d} \right\rangle^2
  = \left\langle h(j) Q^{(2)}_{j,1} \right\rangle^2+O\left( \frac{1}{F^4} \right) \nonumber 
     \\
    &\!=\! \frac{1}{16}-\frac{1}{4}\frac{1}{F^2} +O\left( \frac{1}{F^4} \right).
\end{align}
Finally, one can determine the norm of the true perturbation introduced in Eq. (\ref{h_dd_perp}). Due to orthogonality of $h^\perp(j)$ and $h^\parallel(j)$ one finds
\begin{align}
    ||h^\perp (j)||^2 &= ||h (j)||^2 - ||h^\parallel (j)||^2 =  \frac{1}{4}\frac{1}{F^2} +O\left( \frac{1}{F^4} \right),
\label{norm}
\end{align}
i.e., the squared norm of the true perturbation decays as $ 1/F^2$. It is clear that the strength of the perturbation in the present model is not determined solely by $V$, but it decreases significantly with $F$. It poses a challenge to the finite size numerics, as for sufficiently strong tilts the true perturbation becomes so small that the studied model appears localized.   A similar result has been established numerically for strongly disordered chains ~\cite{krajewski_vidmar_22}. In the latter case $||h^\perp (j)||^2$ decays as $\propto 1/W^2$ where $W$ is the disorder strength. 

\begin{figure}[!ht]
\includegraphics[width=1.0\columnwidth]{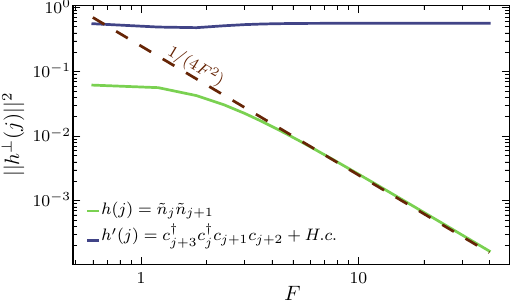}
\caption{
Squared norms of the density of true perturbation defined  in Eqs.(\ref{h_dd_par}) and (\ref{h_dd_perp}) as a function of tilt $F$. Continuous lines show numerical results obtained for $L=18$ at half fillings for two types of the many-body interactions, as indicated in the legend. Dashed line shows analytical result from Eq. (\ref{norm}).
}
\label{fig1}
\end{figure}

In order to estimate the applicability of the $1/F$ expansion for finite tilts, we have also carried out numerical studies of a finite system with
$L=18$ sites. We have first calculated the single particle wave-functions
$\langle j | \alpha \rangle$ and then numerically constructed the occupations of the Stark-Wannier states $Q_{\alpha}$ as well as their products $Q^{(2)}_{\alpha,d}$ for $d=1,2$. 
 Then, the true perturbation $h^\perp(j)$ and its norm are calculated using Eqs.~ (\ref{h_dd_par}) and (\ref{h_dd_perp}). The results are shown in Fig.~\ref{fig1}. In the high-field regime, the numeric results fit very accurately the analytical results from Eq. (\ref{norm}), whereas quite reasonable agreement between both results takes place already for $F \ge 2$.

We stress that such quick decay of the true perturbation with $F$ is not a generic property of two-body interactions. As an example in Fig.~\ref{fig1} we show numerical results for the case when the 
density-density interaction in Eq. (\ref{h_dd}) is replaced by a pair hopping $h(j) \to h'(j) =c^{\dagger}_{j+3} c^{\dagger}_{j}   c_{j+1}   c_{j+2}  +{\rm H.c.}$. In such a case, the true perturbation hardly depends on the tilt, i.e., the strength of the perturbation is controlled solely be $V$.

\subsection{The form of the true perturbation}
\label{subsec:form}

Having the analytical form of the single-particle wave functions, one may explicitly calculate the leading contributions to the true perturbation. 
In this subsection, we expand $H_V^\perp=H_V-H_V^\parallel$ into a power series in $1/F$ up to the second order. 
% \begin{equation}
%     H_V^\perp=\textrm{diag}(H_V^\perp) +H_V^{\perp(1)} + H_V^{\perp(2)},
% \end{equation}
% where $H_V^{\perp(1)}$ and $H_V^{\perp(2)}$ are non-diagonal parts of $H_V^\perp$ linear and quadratic in inverse field consecutively.
To do this, we first inspect the matrix elements of $H_V^\parallel$ 

% \begin{align}
%    \begin{split}
%         H_V^\parallel=&V\sum_{j} h^\parallel(j) \\
%        =&V\sum_{j, \alpha, d}  \left( \mathcal{J}_{j-\alpha}\mathcal{J}_{j+1-(\alpha+d)}-\mathcal{J}_{j+1-\alpha}\mathcal{J}_{j-(\alpha+d)} \right)^2 \\
%        &\qquad\quad \sum_{k,l,m,n}\mathcal{J}_{k-\alpha}\mathcal{J}_{l-\alpha}\mathcal{J}_{m-(\alpha+d)}\mathcal{J}_{n-(\alpha+d)}\\
%        &\qquad\qquad\quad \left(c_k^\dagger c_l -\frac{1}{2}\delta_{kl}\right)\left(c_m^\dagger c_n -\frac{1}{2}\delta_{mn}\right).\\
%    \end{split}
% \end{align}
\begin{widetext}
    \begin{equation}
        H_V^\parallel=V\!\!\sum_{j, \alpha, d}\!\!  4 \left\langle h(j) Q^{(2)}_{\alpha,d} \right\rangle \!\!\!\sum_{k,l,m,n}\!\!\!\mathcal{J}_{k-\alpha}\mathcal{J}_{l-\alpha}\mathcal{J}_{m-(\alpha+d)}\mathcal{J}_{n-(\alpha+d)} \left(c_k^\dagger c_l -\frac{1}{2}\delta_{kl}\right)\left(c_m^\dagger c_n -\frac{1}{2}\delta_{mn}\right).
        \label{hvexp}
    \end{equation}
\end{widetext}
Since $H_V$ is diagonal in the Wannier basis, we first inspect the diagonal part of $H_V^\parallel$ in this basis, that will be denoted as $H_V^{\parallel(d)}$.  We have argued in the preceding subsection that the projection $\left\langle h(j) Q^{(2)}_{\alpha,d} \right\rangle $ in Eq. (\ref{hvexp}) is of the order of $\left(\frac{1}{F}\right)^{o_1}$ with $o_1=2(|j-\alpha|+|j-\alpha+1-d|)$ whereas the order of the four other Bessel functions in Eq. (\ref{hvexp}) is $\left(\frac{1}{F}\right)^{o_2}$ with $o_2=|k-\alpha|+|l-\alpha|+|m-\alpha-d|+|n-\alpha-d|$. To single out all diagonal terms up to the second order in $1/F$ we investigate all cases with $o_1+o_2\leq2$ and denote the corresponding contributions to $H_V^{\parallel(d)}$ as $D_{o_1o_2}$. 

The largest contribution to $H_V^{\parallel(d)}$ corresponds to $o_1=o_2=0$ when $\alpha=j$, $d=1$, $k,l=j$ and $m,n=j+1$. Using the expansion of the Bessel functions from Eq.~(\ref{J_exp}) we obtain
% \begin{equation}
%     D_{00}=V\sum_j \left(1-\frac{3}{2}\left(\frac{2t}{F}\right)^2\right)\left(n_j-\frac{1}{2}\right)\left(n_{j+1}-\frac{1}{2}\right).
% \end{equation}
\begin{equation}
    D_{00}=V\sum_j \left(1-6\frac{1}{F^2}\right)\tilde{n}_j\tilde{n}_{j+1}.
\end{equation}
There are 6 diagonal terms corresponding to 
$o_1=0$ and $o_2=2$ which sum to
% \begin{equation}
% \begin{split}
%          \MoveEqLeft[9]D_{02}=V\sum_j \frac{1}{2}\left(\frac{2t}{F}\right)^2\left[ \left(n_j-\frac{1}{2}\right)\left(n_{j+1}-\frac{1}{2}\right)\right. \\&+\left.\left(n_j-\frac{1}{2}\right)\left(n_{j+2}-\frac{1}{2}\right)\right].
% \end{split}
% \end{equation}
\begin{equation}
         D_{02}=2V\sum_j \frac{1}{F^2}\Big[\tilde{n}_j\tilde{n}_{j+1}+\tilde{n}_j\tilde{n}_{j+2}\Big].
\end{equation}
Finally, for $o_1=2$ and $o_2=0$ one gets two  contributions: $\alpha=j$, $d=2$ or $\alpha=j-1$, $d=2$ with all other indices set accordingly to fulfill $o_2=0$ for which one finds 
% \begin{equation}
%     D_{20}=V\sum_j\frac{1}{2}\left(\frac{2t}{F}\right)^2\left(n_j-\frac{1}{2}\right)\left(n_{j+2}-\frac{1}{2}\right)
% \end{equation}
\begin{equation}
    D_{20}=2V\sum_j\frac{1}{F^2}\tilde{n}_j\tilde{n}_{j+2}.
\end{equation}
% \begin{equation}
% \begin{split}
%          \MoveEqLeft[9.5]\textrm{diag}(H_V^\parallel)=V\sum_j \left(1-\left(\frac{2t}{F}\right)^2\right)\left(n_j-\frac{1}{2}\right)\left(n_{j+1}-\frac{1}{2}\right)\\&+\left(\frac{2t}{F}\right)^2\left(n_j-\frac{1}{2}\right)\left(n_{j+2}-\frac{1}{2}\right).
% \end{split}
% \end{equation}
%\begin{equation}
%\begin{split}
%         \MoveEqLeft[14]\textrm{diag}(H_V^\parallel)=V\sum_j \left(1-4\frac{1}{F^2}\right)\tilde{n}_j\tilde{n}_{j+1}\\&+4\frac{1}{F^2}\tilde{n}_j\tilde{n}_{j+2}.
%\end{split}
%\end{equation}
After summing up all diagonal terms,  \mbox{$H_V^{\parallel(d)}=D_{00}+D_{20}+D_{02}$},
 it is straightforward to calculate also the diagonal part of the true perturbation
% \begin{equation}
% \begin{split}
%          \MoveEqLeft[10.5]\textrm{diag}(H_V^\perp)=V\sum_j \left(\frac{2t}{F}\right)^2\left[\left(n_j-\frac{1}{2}\right)\left(n_{j+1}-\frac{1}{2}\right)\right.\\&-\left.\left(n_j-\frac{1}{2}\right)\left(n_{j+2}-\frac{1}{2}\right)\right].
% \end{split}
% \end{equation}
\begin{equation}
        H_V^{\perp(d)}=H_V-H_V^{\parallel(d)}=\frac{4V}{F^2} \sum_j  \tilde{n}_j \left(n_{j+1}-n_{j+2}\right).
        \label{hvpd}
\end{equation}
Diagonal part of the true perturbation, $H_V^\perp$, decays quadratically with field and consists of the nearest- and the next-nearest neighbor density-density interaction. 

%Interaction of this form is commonly used to study tilted chains where the latter term is used to delocalize the system at $F=0$ [cite].

Next, we focus on the off-diagonal terms of the true perturbation, starting with the contribution that is of the order of $\frac{1}{F}$. Linear terms are obtained from Eq.~(\ref{hvexp}) only for $o_1=0$ and $o_2=1$, given that $o_1$ has to be an even number.
The former condition yields $\alpha=j$ and $d=1$ while the latter condition, $|k-j|+|l-j|+|m-j-1|+|n-j-1|=1$,  means that only one of these four summands is nonzero.  From these constraints one obtains four terms describing correlated hoppings which sum up to the following expression
\begin{equation}
    H_V^{\perp(1)}=V\sum_j\frac{1}{F}(n_{j+3}-n_{j})\left(c_{j+1}^\dagger c_{j+2}+H.c.\right).
            \label{hvp1}
\end{equation}
This part of the true perturbation is odd under the inversion transformation $c_{j} \to c_{-j}$ but is invariant when the inversion is combined with $F \to -F$.   

Finally, we calculate the quadratic off-diagonal terms of the true perturbation. They are obtained for $o_1=0$ and $o_2=2$ which yields $\alpha=j$, $d=1$ and $|k-j|+|l-j|+|m-j-1|+|n-j-1|=2$. This gives one several terms which can be compactly written as 
% \begin{equation}
% \begin{split}
%          \MoveEqLeft[0.5]H_V^{\perp(2)}=V\sum_j \frac{1}{8}\left(\frac{2t}{F}\right)^2\Bigg[(2n_{j+2}-n_j-n_{j+4})\\&\qquad\qquad\qquad\qquad\qquad\qquad\quad\left(c_{j+1}^\dagger c_{j+3}+H.c.\right)\\&+2\left(c_{j}^\dagger c_{j+3}^\dagger c_{j+2}c_{j+1} + c_{j}^\dagger c_{j+2}^\dagger c_{j+3}c_{j+1} + H.c.\right)\Bigg].
% \end{split}
% \end{equation}
\begin{widetext}
\begin{equation}
         H_V^{\perp(2)}=\frac{V}{2}\sum_j \frac{1}{F^2}\Bigg[(2n_{j+2}-n_j-n_{j+4})\left(c_{j+1}^\dagger c_{j+3}+H.c.\right)+2\left(c_{j}^\dagger c_{j+3}^\dagger c_{j+2}c_{j+1} + c_{j}^\dagger c_{j+2}^\dagger c_{j+3}c_{j+1} + H.c.\right)\Bigg].
                     \label{hvp2}
\end{equation}
\end{widetext}

Having calculated the explicit form of the true perturbation up to the second order in  the inverse field $H_V^{\perp}=H_V^{\perp(d)}+H_V^{\perp(1)}+H_V^{\perp(2)}$, we may check how each contribution to the true perturbation influences the dynamics of the tilted model. 
%and compare it to the dynamics of commonly used model that also emerged in the diagonal of true perturbation (up to the sign).

\section{Charge dynamics}
\label{sec:dynamics}

Analytical calculations in the preceding section show that the true perturbation contains three types of the two-body interaction:
density-density coupling in Eq.~(\ref{hvpd}), 
 correlated hoppings  in  
 Eqs.~(\ref{hvp1}) and (\ref{hvp2}) as well as the pair-hopping
 interactions in Eq.~(\ref{hvp2}). The corresponding coupling strengths decay either as $1/F$ or as $1/F^2$ and for large $F$ the perturbations become too small to be studied numerically in finite systems. In order to overcome this problem and to compare their influence on the charge dynamics, we rescale their strengths and study them separately.  
 
 In order to reduce the finite-size effects, we numerically investigate charge dynamics in a model which is equivalent to the tilted chain but allows for periodic boundary conditions. In this model the field $F$ is introduced via time-dependent flux 
\begin{equation}
    H_F=\sum_j\left( e^{-iFt}c_j^\dagger c_{j+1} + e^{iFt} c_{j+1}^\dagger c_{j}\right) + H'.
    \label{Hf}
\end{equation}
We choose various forms of the interaction term $H'$ which occur in the true perturbation. As a reference we take a density-density interaction on the nearest-neighbor and the next-nearest-neighbor sites
\begin{equation} H'_1=V'_1\sum_j\Big[\tilde{n}_j\tilde{n}_{j+1}+\tilde{n}_j\tilde{n}_{j+2}\Big],
\end{equation}
as well as the similar term which represents the diagonal part  of the true perturbation, see Eq.~(\ref{hvpd})
\begin{equation}
         H'_2=V'_2\sum_j\Big[\tilde{n}_j\tilde{n}_{j+1}-\tilde{n}_j\tilde{n}_{j+2}\Big].
\end{equation}
 We  consider also both off-diagonal parts of $H_V^\perp$, see Eqs.~(\ref{hvp1}) and (\ref{hvp2}), rewritten in a form appropriate for the time-dependent flux
 \begin{equation}
     H'_3=V'_3\sum_j (n_{j+3}-n_{j})\left(e^{-iFt}c_{j+1}^\dagger c_{j+2}+H.c.\right),
 \end{equation}
% \begin{equation}
% \begin{split}
%          \MoveEqLeft[3]H'_4=V'_4\sum_j \Bigg[(2n_{j+2}-n_j-n_{j+4})\\&\qquad\qquad\qquad\qquad\qquad\left(e^{-2iFt}c_{j+1}^\dagger c_{j+3}+H.c.\right)\\+2\Big(&c_{j}^\dagger c_{j+3}^\dagger c_{j+2}c_{j+1} + e^{-2iFt}c_{j}^\dagger c_{j+2}^\dagger c_{j+3}c_{j+1} + H.c.\Big)\Bigg],
% \end{split}
% \end{equation}
\begin{widetext}
\begin{equation}
        H'_4=V'_4\sum_j \Bigg[(2n_{j+2}-n_j-n_{j+4})\left(e^{-2iFt}c_{j+1}^\dagger c_{j+3}+H.c.\right)\\+2\Big(c_{j}^\dagger c_{j+3}^\dagger c_{j+2}c_{j+1} + e^{-2iFt}c_{j}^\dagger c_{j+2}^\dagger c_{j+3}c_{j+1} + H.c.\Big)\Bigg],
\end{equation}
\end{widetext}
alongside a symmetric pair-hopping term from $H_V^\perp(2)$ being also the simplest effective model for a tilted chain
\begin{equation}
    H'_5=V'_5\sum_j \left( c_{j}^\dagger c_{j+3}^\dagger c_{j+2}c_{j+1} + H.c. \right).
\end{equation}
In numerical calculations we set $V'_1=2$ and then we determine the remaining potentials $V'_2$, $V'_3$, $V'_4$ and $V'_5$ in such a way that the norms of all interaction terms are equal 
\begin{equation}
||H'_1||^2=\langle H'_1 H'_1\rangle =||H'_2||^2=...=||H'_5||.    
\end{equation}

The charge dynamics is studied numerically using the same technique as in Ref. \cite{nandy2024}.
For the sake of completeness,  we briefly recall the main steps. 
Initially, we prepare a thermal state for the Hamiltonian $H_{t<0}=H_{F=0}(t)+\sum_j \cos(q\; j)\tilde{n}_j$. To this end, we utilize the microcanonical Lanczos method \cite{long03,prelovsek11,herbrych22}. We use a finite but sufficiently high temperature $kT=10$, so that the density modulation induced by the term $\sum_j \cos(q \;j)\tilde{n}_j$ remains within the linear response regime, i.e. the modulation has a form of a plain wave with wave-vector $q$ and the amplitude $A_0$, $\langle\tilde{n}_j\rangle = A_0 \cos(qj)$.  
 Then we quench the field $F$ at $t=0$ and observe the evolution of the amplitude of the modulation, $A_t$. This amplitude is obtained from the discrete Fourier transform performed on $L$-dimensional vector $[\langle\tilde{n}_1\rangle,...,\langle\tilde{n}_L\rangle]$, whereas the evolution is calculated using the Lanczos propagation method \cite{lantime1,lantime2}.

\begin{figure}[!ht]
\includegraphics[width=1.0\columnwidth]{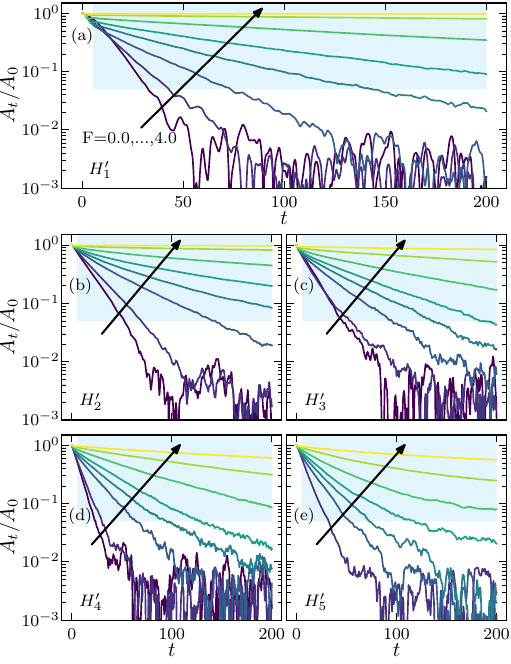}
\caption{
Time evolution of the normalized amplitude $A_t$ of density modulation with wave vector $q=2\pi/L$ at $L=24$ for Hamiltonian $H_F(t)$ defined in Eq.~(\ref{Hf}) and $F=0.0,0.4,0.8,1.0,1.2,1.6,2.4,4.0$. The arrows represent the increasing $F$. Panels (a)-(e) contain results for $H'_1$-$H'_5$, respectiely. Shaded areas depict the range where the data were fitted by the exponential function $A_t/A_0=\exp(-\Gamma t)$.
}
\label{fig2}
\end{figure}

The results for the evolution of normalized wave amplitude $A_t/A_0$ are presented in Fig. \ref{fig2} separately for each interaction $H'=H'_1,...,H'_5$. We use $L=24$ and the smallest wave-vectors $q=2\pi/L$. For the accessible system sizes, we
cannot reliably determine how the decay rate $\Gamma$ depends on  $q$. However, we expect the transport to be either diffusive with $\Gamma\propto q^2$ or subdiffusive $\Gamma\propto q^4$ and in both cases the amplitude of the density modulation decays exponentially in time \cite{nandy2024}, $A_t=A_0 \exp(-\Gamma t)$, as it is also visible from numerical results in Fig. \ref{fig2}.    

\begin{figure}[!ht]
\includegraphics[width=1.0\columnwidth]{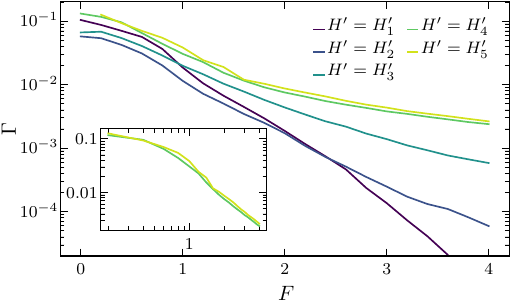}
\caption{
Decay rate of normalized amplitude of charge density wave with wave vector $q=2\pi/L$ obtained from exponential fits of shaded area in Fig.~\ref{fig2}. The inset shows  results for $H'_4$ and $H'_5$ on log-log scale.
}
\label{fig3}
\end{figure}

 The relaxation rates, $\Gamma$, have been obtained from fitting results for $A_t$. We have used results for sufficiently large amplitude,
 $A_t \ge 0.05 A_0$, (marked by shaded area in Fig.~\ref{fig2}) 
 for which  the deviations from the exponential decay are rather inessential.  The relaxation rates are shown in Fig.~\ref{fig3}.  They are obtained for various forms of $H'$ but with equal normalizations. In case of weak tilts, $F\lesssim 0.5$,  the relaxation rates are of the same order of magnitude. However, upon increasing $F$ the differences between various interactions become very pronounced. The smallest relaxation rate and the slowest charge dynamics take place for the density-density interactions,
$H'_1$ and $H'_2$, which are mostly studied in the context of Stark MBL. The relaxation rate is an order of magnitude larger for the case of correlated hopping $H'_3$. However, the fastest dynamics is observed for either $H'_4$ or $H'_5$, whereby the latter two interactions lead to almost identical relaxation rates. The decay of the relaxation rates for $H'_4$ and $H'_5$ seems to be a power law (see the inset in Fig.~\ref{fig3}), as oppose to a clear exponential decay for $H'_1$. However, this observation should be verified for larger span of $F$ that is not accessible to our numerical calculations. Nevertheless,  Fig.~\ref{fig3} shows that the pair hopping interaction described by $H'_5$ is an efficient source of the charge dynamics in strongly tilted chains and the resulting relaxation rate can be two orders of magnitude larger than in the case of density-density interaction. 

It is interesting that $H'_5$, when studied alone as the entire Hamiltonian, strictly conserves the dipole moment. It has been considered as an effective model \cite{kohlert2023,scherg2021,moudgalya2020} for the strong-tilted systems that exhibit the Hilbert space fragmentation \cite{pai2019,khemani2020,moudgalya2022,francica2023,brighi2023,will2023,sala2020,lydzba2024}.

\section{Summary}
\label{sec:summary}
The fate of the single particle localization in the presence of the many-body interactions has recently attracted a significant interest. The cases of  the single-particle Anderson insulator and the Stark localization have been studied in the context of MBL and Stark MBL, respectively. One usually considers the simplest many-body interaction, $H_V$, being the nearest-neighbor density-density coupling. One may formally single out a part of $H_V$ which does not commute with the one-particle Hamiltonian and represents the true perturbation to the single-particle localization. Such procedure has been carried out numerically for disordered systems \cite{krajewski_vidmar_22} whereas in present work we derive an explicit form of the true perturbation for chains tiled by the field $F$ taking into account terms up to the order $1/F^2$. The main contributions to the true perturbations describe the nearest-neighbor correlated hopping and the pair hopping interactions. Finally, we have shown that the charge dynamics in the Stark chain significantly depends on the choice of the two-body interaction. When the density-density coupling is replaced by the pair-hopping interaction of the same magnitude then the charge dynamics may speed up by a few orders of magnitude. Therefore, the slow charge dynamics in the mostly studied model with the density-density interaction originates, at least partially, from the smallness of the true perturbation and seems not to be generic for other two-body interactions. In this respect, the Stark MBL systems closely resemble the disordered MBL chains.

\acknowledgments 
We acknowledge support from the National Science Centre, Poland via Project No. 2020/37/B/ST3/00020. The numerical calculations were partly carried out at the facilities of the Wrocław Centre for Networking and Supercomputing.
 
\bibliographystyle{biblev1}
\bibliography{ref_mbl}
 
\end{document}